\documentclass[aps,prd,amsmath,floats,floatfix, twocolumn,
superscriptaddress,nofootinbib,showpacs,longbibliography]{revtex4-1}

\usepackage[T1]{fontenc}
\usepackage[utf8]{inputenc}
\usepackage{lmodern}
\usepackage{physics}
\usepackage{verbatim}

\usepackage[dvipsnames, usenames]{xcolor}
\definecolor{linkcolor}{rgb}{0.0,0.3,0.5}
\usepackage[hypertexnames=false, unicode, colorlinks=true, linkcolor=linkcolor,
citecolor=linkcolor, filecolor=linkcolor,urlcolor=linkcolor,
pdfusetitle]{hyperref}

\usepackage[all]{hypcap}
\usepackage{graphicx}
\usepackage{xspace}
\usepackage{amssymb}
\usepackage[normalem]{ulem} 
\usepackage{bm} 

\usepackage{microtype}

\usepackage[english]{babel}
\usepackage{blindtext}

\usepackage{float}

\graphicspath{%
  {figs/}%
}

\DeclareMathAlphabet{\mathpzc}{OT1}{pzc}{m}{it}

\newcommand{\TypeI}{Type-I}
\newcommand{\TypeII}{Type-II}

\newcommand{\qrec}{\ensuremath{\mathcal{Q}_\mathrm{rec}}}
\newcommand{\AEI}{\affiliation{Max Planck Institute for Gravitational Physics (Albert Einstein Institute), Am M\"uhlenberg 1, Potsdam 14476, Germany}}

\def\msun{\,{\rm M_\odot}}

\def\be{\begin{equation}}
\def\ee{\end{equation}}
\def\bea{\begin{eqnarray}}
\def\eea{\end{eqnarray}}
\newcommand{\bes}{\begin{subequations}}
	\newcommand{\ees}{\end{subequations}}

\allowdisplaybreaks

\graphicspath{{./Figures/}}

\begin{document}
	
	\pagenumbering{arabic}
	
	\title{Detection and parameter estimation challenges of Type-II lensed binary black hole signals}
	\author{Aditya Vijaykumar}
	\affiliation{International Centre for Theoretical Sciences, Tata Institute of Fundamental Research, Bangalore 560089, India}
	\author{Ajit Kumar Mehta}\AEI
	\author{Apratim Ganguly}
	\affiliation{Inter-University Centre for Astronomy and Astrophysics (IUCAA), Post Bag 4, Ganeshkhind, Pune 411 007, India}
	\affiliation{International Centre for Theoretical Sciences, Tata Institute of Fundamental Research, Bangalore 560089, India}
	
	\date{\today}

	\begin{abstract}

	Strong lensing of {gravitational-wave signals} can produce three types of images, denoted as Type-I, Type-II and Type-III, corresponding to the minima, saddle and maxima of the lensing potential of the lensed images. Type-II images, in particular, receive a non-trivial phase shift of $\pi/2$. This phase shift can introduce additional distortions in the strains produced by the Type-II image of the binary black hole signals depending on the morphology of the signals, e.g., when they have contributions from higher harmonics, precession, eccentricity, etc. {The probability of observing Type-II images is nearly the same as that of strong lensing itself, and thus these signals are likely to be observed in the near future.} In this work, we investigate the potential applicability of these distortions in helping identify Type-II signals from a single detection and the systematic biases that could arise in the inference of parameters if they are {recovered with gravitational-wave templates that do not take the distortion into account}. {We show that the lensing distortions will allow us to confidently identify the Type-II images for highly inclined binaries: at network signal-to-noise ratio (SNR) $\rho=20(50)$, individual Type-II images should be identifiable with ln Bayes factor $\ln \mathcal{B} > 2$ for inclinations $ \iota > 5 \pi/12  (\pi/3) $. }Furthermore, based on the trends we observe in these results, we predict that, at high SNRs ($\gtrsim 100$), individual Type-II images would be identifiable even when the inclination angle is much lower ($\sim \pi/6$). We then show that neglecting physical effects arising from these identifiable Type-II images can significantly bias the estimates of source parameters (such as sky location, distance, inclination, etc.). {Thus, in the future, using templates that take into account the lensing deformation would be necessary to extract source parameters from Type-II lensed signals.}
	
	\end{abstract}
	
	
	\maketitle

	\section{Introduction}
	
	The sensitivities of the LIGO \& Virgo gravitational-wave (GW) detectors \cite{2015CQGra..32g4001L,2015CQGra..32b4001A} are constantly improving. So far, they have already confidently detected $90$ compact binary mergers events \cite{LIGOScientific:2021djp} and are expected to detect hundreds of such mergers in the upcoming observing runs \cite{KAGRA:2013rdx}. Accurate extraction of source parameters from these mergers is essential to interpret them in an astrophysical and cosmological context. Understanding and modeling physical effects due to the propagation of the wave between the source and the detector is significant in this regard. Ignoring these effects in the templates used to search and estimate parameters of a GW signal could cause problems in explaining or interpreting the underlying signals.
	
	One such propagation effect is gravitational lensing. GWs get lensed when the signals encounter mass inhomogeneities during their journey to the detectors. If the intervening object is sufficiently massive, it can produce multiple images of the source through strong lensing \cite{Dai:2016igl, Ng:2017yiu}. The multiple images would differ in their magnifications and arrival times at the observer. The probability (quantified by the optical depth) for strong lensing is, however, small, e.g., $\sim 10^{-3} - 10^{-4}$ for galaxy lenses \cite{Wang:2021kzt}, mainly because it requires a strong alignment between the source, lens and the observer. Searches in GW data have not detected signatures of strong lensing so far \cite{Hannuksela:2019kle, LIGOScientific:2021izm}. {Nevertheless, at the design sensitivities of Advanced LIGO \& Virgo \cite{LIGOScientific:2014pky, VIRGO:2014yos}, $1.3^{+0.6}_{-0.4} - 1.7^{+0.9}_{-0.6}$ detections of binary black holes (BBH) lensed by galaxies are expected per year \cite{Wierda:2021upe}. } These prospects will only be enhanced with the addition of KAGRA \cite{Aso:2013eba, KAGRA:2020tym} and LIGO-India \cite{LI-Det, Saleem:2021iwi} to the detector network. Furthermore, third-generation GW detectors like Cosmic Explorer (CE) \cite{Reitze:2019iox} and Einstein Telescope (ET) \cite{Sathyaprakash:2012jk} are expected to have an order of magnitude better sensitivity as compared to the current generation of detectors, thus potentially observing hundreds of thousands of mergers every year. This suggests an exciting time for doing science with strongly lensed GWs in future~\cite{Sereno:2011ty, Ding:2015uha, Baker:2016reh, Hannuksela:2020xor, Goyal:2020bkm, Xu:2021bfn}.
	
	In the geometric-optics limit, the lens equation picks solutions at extremal points of the time-delay (arrival time with respect to a reference time) surface over a family of trajectories. The images thus produced due to strong lensing can be of different types corresponding to the minima, saddle point and the maxima of the time-delay, called Type-I, Type-II and Type-III images, respectively. Type-I and Type-III images have positive parity (i.e., positive magnification), while Type-II images have negative parity. In addition, Type-II images receive a $\pi/2$ phase shift, while Type-I and Type-III images receive a phase shift of $0$ and $\pi$, respectively \cite{Dai:2017huk}. Thus, unlensed signals with an overall magnification can mimic Type-I and Type-III images. On the other hand, Type-II images get distorted compared to their unlensed counterparts due to the non-trivial phase shift of $\pi/2$. In the cases when the GW radiation is mainly composed of the dominant harmonics ($2, \pm 2$ modes), these distortions can be mimicked by Type-I or the unlensed GW signals by adjusting their parameters, such as the coalescence phase and/or polarization angle \cite{Ezquiaga:2020gdt}. However, when the signal has significant contributions from higher harmonics, precession, or eccentricity, these distortions may not be reproduced by the Type-I signals \cite{Ezquiaga:2020gdt}, as also discussed in Sec.~\ref{sec:setup} here. In such cases, the current strategies that LIGO-Virgo follows, i.e., analyzing a detected signal under the hypothesis that the signal is unlensed, may lead to significant biases in the inferences of the parameters. The standard LIGO-Virgo search pipelines could also miss the signals \cite{Ezquiaga:2020gdt, Wang:2021kzt}.  
	
    Under the singular isothermal ellipse (SIE) approximation~\footnote{A more realistic approximation than the singular isothermal sphere (SIS).} for the galaxy lenses \cite{SIE_model}, it has been shown that the probability of seeing multiple images without a Type-II image being one of them is less than $0.01\%$ \cite{Wang:2021kzt}. {Hence, the probability of observing Type-II images is practically the same as that of strong lensing.} Moreover, it was also shown that more than $90\%$ of sources with multiple images would have a Type-II image as the second brightest image \cite{Wang:2021kzt}. Thus, if a strongly lensed event pair is detected, there is a high chance that a Type-II image would be present.

    In this work, we thoroughly investigate the systematic effects that could arise when a Type-II lensed (BBH) signal is recovered with the Type-I/unlensed templates. We first establish the cases where a single-event-based confident identification of the Type-II nature of a signal could be possible. This is done by choosing a threshold for the ln Bayes factor between the two hypotheses $\textbf{A}$ and $\textbf{B}$, where  $\textbf{A}$ represents the hypothesis where we inject a Type-II signal and recover with Type-II templates, and $\textbf{B}$ represents the hypothesis when the Type-II injected signal is recovered with the Type-I template. We claim a Type-II signal is identified when the ln Bayes factor $\ln \mathcal{B}$ $\geq 2$. We study the effect of strong lensing on a broad set of injections at total SNR of 20 and 50, using the noise-spectra of the LIGO-Virgo design sensitivities. We show that even at modest values of SNR, the Type-II nature of the image can be inferred for moderate to high values of the inclination angle. {We also show that non-inclusion of the $ \pi/2$ phase shift in the parameter estimation templates for such signals could lead to significant systematic errors, rendering the recovered posteriors inconsistent with the true values.} In other words, the parameters of the signals could be wrongly inferred if Type-II templates are not used during the parameter estimation (PE). This could, in turn, bias the astrophysical interpretation of the source, e.g., a lighter BH could be wrongly attributed to higher mass, thus causing difficulties in explaining their formation mechanisms. 
	
    One strategy to look for the strong lensing signatures in the data is by analyzing pairs of events with consistent sky locations and chirp masses \cite{Haris:2018vmn, Goyal:2021hxv}~\footnote{Ref. \cite{Caliskan:2022wbh} quantifies the false alarm probabilities associated with this strategy and the difficulties that arise in the identification of lensed events as a result.}. This is motivated by the belief that strong lensing will not affect the frequency profile of the observed signal. This is indeed true for Type-I and Type-III images, and in some cases, even for the Type-II images if they are quasicircular and contain only the dominant (22) mode. {In general, though, one would need to perform a joint analysis,  where a pair of events are simultaneously analyzed  \cite{Liu:2020par, Lo:2021nae, Janquart:2021qov}. However, as we discussed before, in a realistic scenario, if there is a lensed event pair in the data, one of them would likely be a Type-II image. Then, given that we would not know \textit{a priori} the parameters of the underlying unlensed GW signal, we may not expect that there will be consistency in the parameters (such as the chirp mass and the sky location parameters) extracted using the templates of the unlensed signals. Thus, in principle, one would need to do a joint-PE search over the full data without any prior belief. This may become a difficult task for such techniques, given their computational costs. On the other hand, doing a full search over data for just the individual Type-II images should be relatively easy, and if detected, they can help narrow down the search for its corresponding pair since we would already know the parameters of the underlying unlensed signal. Moreover, when the lensed pair can not be detected due to detector downtime or data quality issues, lensing signatures can be obtained from just one image and further used for downstream analyses such as estimating lensing rates.}
	
	We note that a similar study has been performed in Ref. \cite{Wang:2021kzt}. However, the study therein is based on the analytical approximation of the Bayes factor, emphasizing the identification of the Type-II signals. They also do not explicitly consider the detector response of the GW signals, avoiding this by assuming that both polarizations are measurable. In our work, we perform a full parameter estimation study, emphasizing both the identification of the Type-II signals and the biases in the inference of the binary parameters. Thus, in this sense, our work builds upon their work along with this particular goal. We also note that Ref. \cite{Janquart:2021nus} has performed a complementary analysis, showing that including higher-order modes in joint-PE allows identifying of individual image types in a lensed pair.   
	
	This paper is organized as follows: in Sec.~\ref{sec:setup} we explain the effect of the strong lensing on GWs and describe the method followed in this work. In Sec.~\ref{sec:results}, we discuss the results based on a wide set of simulations tuned to our interests. Sec.~\ref{sec:conclusion} presents the conclusion and future work. 
	
	\begin{figure*}[htb!]
		\includegraphics[width=0.95\textwidth]{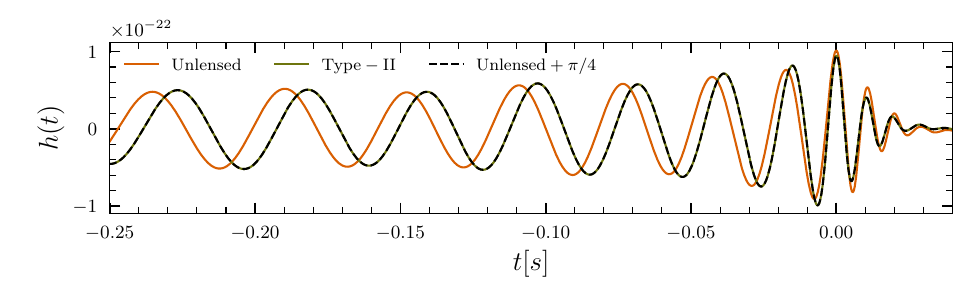}
		\includegraphics[width=0.95\textwidth]{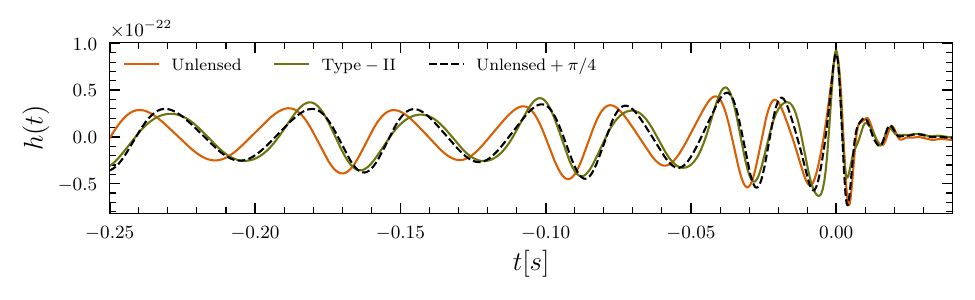}
		\caption{{The unlensed signal, the Type-II signal, and the unlensed signal with a shift in the coalescence phase $\phi_c$ by $\pi/4$ for the parameters $M = 100 M_{\odot}$, $q=7$, $\iota=0$ (top panel) and $\iota=\pi/3$ (bottom panel). }For the zero inclination (top panel), i.e., when mainly 22 mode contributes, the GR signal reproduces the Type-II image signal very well by adjusting its coalescence phase while for high inclination (bottom), Type-II signal is not reproduced by the GR signal fully, i.e., there are additional distortions left.  }
		\label{fig:lensed_waveforms}
	\end{figure*}
	
	\section{Set up}
	\label{sec:setup}
	\subsection{Theory}
	
	The problem of lensing of GWs has been looked at in-depth already \cite{PhysRevD.59.083001}. In geometric optics limit (i.e., the short-wavelength regime, valid for, e.g., lensing of GWs by galaxies), the effect of lensing on GWs can be derived from the Kirchhoff diffraction integral~\footnote{The Kirchhoff diffraction integral describes the generic lensing phenomenon.} by choosing points on the image plane corresponding to the extremal points of the time-delay \cite{1986ApJ...310..568B}. Since the extremal points can be either the minimum, saddle, or the maximum, we can have three types of images: Type-I, Type-II, and Type-III, respectively. In this limit, the Kirchhoff integral reduces to the Gaussian integral around these extremal points and the result is the following amplification factor for $j^{\rm{th}}$ image,
	
	\begin{equation}
	F_{j}(f) = \sqrt{\mu_j}\exp[ 2\pi i f \delta t_j- \text{sign}(f)i n_j\pi]
	\end{equation}
	where $f$ is the observed GW frequency, $\mu_j$ and $\delta t_j$ represent the magnification and time delay of the $j^{\rm{th}}$ image, and the extra phase shift $\Delta \phi_{j}= \text{sign}(f) n_j\pi$ results from the integration of the complex Gaussian function, where $n_j$, called the Morse index, takes the values $0$, $1/2$ or $1$ for Type-I, II or III images, respectively. The lensed GW signal is then given by convolving the unlensed signal with the amplification factor,  
	
	\begin{equation}
	\Tilde{h}_{+,\times,\, j}^{L} (f) = F_{j}(f) \ \Tilde{h}_{+,\times}(f), 
	\label{eq:amp_fact}
	\end{equation}
	As we can see, the Type-I image receives just a linear phase compared to the unlensed signal. This linear phase is practically un-important for all of our purposes since this can not be measured by the PE. The Type-III image receives an additional overall negative sign but otherwise is again the same as the unlensed signal. The Type-II image, on the other hand,  depending on the sign of the frequencies, can receive $+\pi/2$ or $-\pi/2$ phase shifts. As a consequence, the time-domain lensed signal is given by the Hilbert transformation of Eq.~\eqref{eq:amp_fact} rather than the simple inverse Fourier transformation of the unlensed signal,
	
	\begin{equation}
	h_{+,\times}^{L,\mathrm{II}} (t) = - \int_{-\infty}^{\infty} df \mathrm{sign} (f) i e^{-i2\pi f t}  \Tilde{h}_{+,\times}(f)
	\label{eq:time_dom_lensed_wav}
	\end{equation}
	
	For our purpose in this work, we directly use Eq.~\eqref{eq:amp_fact} for the lensed template construction because the GW data analysis is done in the frequency domain itself. Nevertheless, Eq.~\eqref{eq:time_dom_lensed_wav} serves a better purpose for the visual inspections of the lensing effects and thus, we use it just for the sake of demonstrations. Also, the standard GW data analysis employs only the positive frequencies because, for a real signal, the negative frequencies do not provide any extra information and hence only the $+ \pi/2$ phase shift is relevant for us. We now discuss the implications of the $\pi/2$ phase shift for the Type-II images with different structures. 
	
	A general GW can have different harmonic contents. One way to visualize it is via spherical mode decomposition,
	\begin{equation}
	h_{+}(t) -i h_{\times}(t) = \sum_{\ell \geq 2}\sum_{m=-\ell}^{\ell} h_{\ell m}(t; \bm{\lambda}) _{-2}Y_{\ell m}(\iota, \phi_{c})
	\label{eq:mode_decomp}
	\end{equation}
	where $_{-2}Y_{\ell m}$ \cite{Pan:2011gk} denotes the spin-2 weighted spherical harmonics which are function of the inclination angle $\iota$~\footnote{The angle between the total angular momentum $\vec{\boldsymbol{J}}$ of the binary and the observer.} and the coalescence phase $\phi_c$. The spherical harmonics basically separate the angular part of the GW radiation from its radial part and thus the modes $h_{\ell m}(t)$ depend just on the intrinsic parameters ($\bm{\lambda}$) of the binary, e.g., component masses ($m_{1,2}$) and spins ($\chi_{1,2}$) if the binary is circular. For non-precessing BBH binaries in circular motion, the $m<0$ modes are related to the $m>0$ due to the reflection symmetry of the GW radiation about the binary plane,
	\begin{equation}
	h_{\ell, -m} = (-1)^{\ell} h^{*}_{\ell m}
	\end{equation}
	This helps simplify Eq.~\eqref{eq:mode_decomp} in terms of only the $m>0$ modes and thus throughout this paper we use $m>0$ modes to denote the full mode contents. On a bit further simplification of Eq.~\eqref{eq:mode_decomp} writing $h_{\ell m}= A_{\ell m}e^{i \phi_{\ell m}}$, one would obtain (see Appendix C of \cite{Mehta:2017jpq}),
	
	\begin{align}
	h_{+}(t) &=  \sum_{\ell \geq 2}\sum_{m\geq 0} f_{\ell m}^{+}(\iota) A_{\ell m}(t) \cos[\phi_{\ell m}(t) + m \phi_c] \label{eq:plus}\\
	h_{\times}(t) &=  \sum_{\ell \geq 2}\sum_{m\geq 0} f_{\ell m}^{\times}(\iota) A_{\ell m}(t) \sin[\phi_{\ell m}(t) + m \phi_c]
	\label{eq:cross}
	\end{align}
	where $f_{\ell m}^{+}(\iota)$ and $f_{\ell m}^{\times}(\iota)$ are functions of the inclination angle encoding the magnitude of the spherical harmonics $_{-2}Y_{\ell m} (\iota)$. Their explicit expressions are not relevant for the discussion here. The relative contribution of the modes $\ell m$ to the polarizations $h_{+,\times}$ depend on the specific choice of the parameter $\bm{\lambda}$ and the inclination angle ($\iota$). For example, for non-precessing BBH binaries with even the moderate mass ratio $q=m_1/m_2 \geq 4$ the contribution of higher order modes (HMs) becomes significant, neglecting which can have consequences for detection and parameter estimation, e.g., the systematic bias, loss in the detection volume, etc.~\cite{PhysRevD.93.084019,PhysRevD.90.124004, Capano:2013raa}. 
	
	When a GW signal reaches the detectors, the strain induced on a particular detector is given by,
	
	\begin{equation}
	h = F_{+}(\theta, \phi, \psi) h_{+}(t) + F_{\times}(\theta, \phi, \psi) h_{\times}(t)
	\label{eq:strain}
	\end{equation}
	where $F_{+}$ and $F_{\times}$ are called the antenna pattern functions, which arise as a result of the detector's response to the GW signal. They represent the angular sensitivity of that detector and hence is a function of the sky angles $\theta$ and $\phi$ in the detector frame; these parameters denote the location of the binary in the sky. The angle $\psi$ denotes the freedom in fixing the binary plane with respect to the detector plane and is called the polarization angle. The explicit expressions of the antenna pattern functions for the LIGO \& Virgo detectors are\footnote{{These expressions hold when the sky angles are specified in the detector frame.}},
	
	\begin{align}
	F_{+} = &\dfrac{1}{2} \big[1 + \cos^2(\theta)\big ]\cos(2\phi)\cos(2\psi) \nonumber \\
	&-\cos(\theta)\sin(2\phi)\sin(2\psi),  \\
	F_{\times} = & \dfrac{1}{2}\big[1 + \cos^2(\theta)\big ] \cos(2\phi)\cos(2\psi) \nonumber \\
	&+\cos(\theta)\sin(2\phi)\cos(2\psi) 
	\end{align}
	
	By combining Eqs.~\eqref{eq:plus}-\eqref{eq:cross} and  Eq. \eqref{eq:strain} one would obtain, 
	
	\begin{equation}
	h = \sum_{\ell \geq 2}\sum_{m\geq 0} \mathcal{A}_{\ell m} (t) \cos[\phi_{\ell m}(t) + m\phi_c - \chi_{\ell m}]
	\label{eq:final_strain}
	\end{equation}
	
	where,
	
	\begin{align}
	\chi_{\ell m} = \tan^{-1}\Bigg( \dfrac{F_{\times}(\theta, \phi, \psi)f^{\times}_{\ell m}(\iota)}{F_{+}(\theta, \phi, \psi) f^{+}_{\ell m}(\iota)}\Bigg)
	\end{align}
	
	\begin{equation}
	\mathcal{A}_{\ell m} (t) = A_{\ell m}(t) |F_{+}|[1 + \tan^2(\chi_{\ell m})]^{1/2} 
	\end{equation}
	Eq.~\eqref{eq:final_strain} is just a bit more simplified version of Eq.~\eqref{eq:strain}. Now by taking the Fourier transform of the above equation and then applying the Hilbert transformation in Eq.~\eqref{eq:time_dom_lensed_wav} one can easily show that the strains produced by the lensed signals in the time-domain are given as follows: 
	
	For Type-I images, there is only an additional magnification factor $\mu_{I}$, i.e.,
	
	\begin{equation}
	h_{\rm{I}} = \sum_{\ell \geq 2}\sum_{m\geq 0} |\mu_{\rm{I}}|^{1/2} \mathcal{A}_{\ell m} (t_{\rm{I}}) \cos[\phi_{\ell m}(t_{\rm{I}}) + m\phi_c - \chi_{\ell m}]
	\label{eq:final_strain_type1}
	\end{equation}
	
	For Type-II images there is an additional $\pi/2$ phase shift,
	
	\begin{equation}
	h_{\rm{II}} = \sum_{\ell \geq 2}\sum_{m\geq 0} |\mu_{\rm{II}}|^{1/2} \mathcal{A}_{\ell m} (t_{\rm{II}}) \cos\Big[\phi_{\ell m}(t_{\rm{II}}) + m\phi_c - \chi_{\ell m} + \pi/2\Big]
	\label{eq:final_strain_type2}
	\end{equation}
	
	while for Type-III images,
	
	\begin{equation}
	h_{\rm{III}} = \sum_{\ell \geq 2}\sum_{m\geq 0} |\mu_{\rm{III}}|^{1/2} \mathcal{A}_{\ell m} (t_{\rm{III}}) \cos[\phi_{\ell m}(t_{\rm{III}}) + m\phi_c - \chi_{\ell m} + \pi]
	\label{eq:final_strain_type3}
	\end{equation}
	where $t_{j} = t + \delta t_{j}$ for image type $j \in \{\mathrm{I, II, III}\}$.
	As expected, the Type-I and Type-III induced strains are just the rescaled versions of the ones caused by the unlensed GW signals. The extra $\pi$ phase shift in the Type-III strains which causes just a global sign flip can be easily mimicked by the polarization angle $\psi$ through $\chi_{\ell m}$ term under the change $\psi \rightarrow \psi + \pi/2$. Thus, for Type-III image signals, the PE with unlensed GW templates would yield a bias in the polarization angle by $\pi/2$. 
	
	\begin{figure*}[t]
		\centering
		\includegraphics[width=\textwidth]{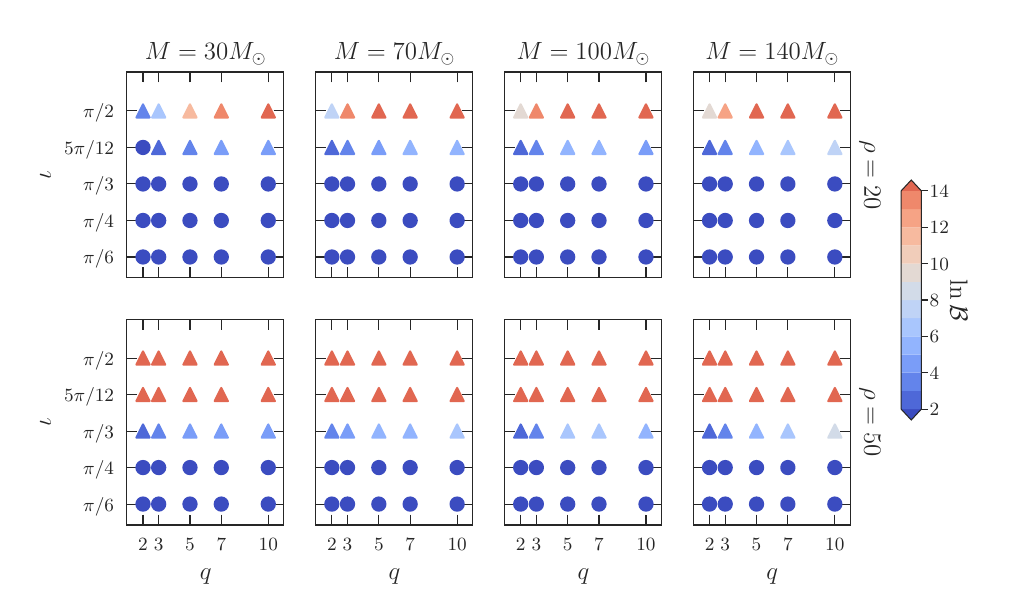}
		\caption{Type-II lensed BBH injections with a network SNRs $\rho= 20$ (upper row) and $\rho=50$ (lower row) shown in the $\iota$-$q$ plane for different total masses. The color bar shows the ln Bayes factor ($\ln \mathcal{B}$) between the Type-II and the Type-I recoveries. Injections which cross our threshold of $\ln \mathcal{B} \geq 2$ are plotted using triangles (i.e., distinguishable image types) while the others are plotted using circles.}\label{fig:Bayes_Factor}
	\end{figure*}
	
	The Type-II induced strain is interesting since the $\pi/2$ phase shift can not be easily absorbed into other known parameters unless the signal is quasicircular, non-precessing and comprised of just the dominant $\ell m =22$ mode, in which case, the coalescence angle $\phi_c$ can easily absorb it by adjusting itself, $\phi_c \rightarrow \phi_c + \pi/4$, as should be clear from Eq.~\eqref{eq:final_strain_type2} when $m=2$. One can note, however, that the polarization angle $\psi$ can also try to adjust itself such that $\Delta \chi_{\ell m} = -\pi/2$ and thus, it is not always the case that only $\phi_c$ will get biased during the PE as we will show later. When HMs also start contributing to the strain, the different $\ell m$ modes would require different shifts in the $\phi_c$, i.e.,
	
	\begin{equation}
	\phi_c \rightarrow \phi_c + \dfrac{\pi}{2m}
	\end{equation}
	in order to mimic the $\pi/2$ lensing phase shift. Thus, the strain produced by Type-II image signals would look distorted compared to its unlensed counterpart. Fig.~\ref{fig:lensed_waveforms} (bottom panel) shows an illustrative example of this case with the binary parameters $M=100 M_{\odot}$, $q=7$ and $\iota = \pi/3$. Since the inclination and the mass ratio are high here, HMs also contribute significantly, and thus there are additional distortions that are not mimicked by the change in the coalescence phase. However, we also show the case with the zero inclination for the same binary configuration in the top panel. For very small inclinations, the GW signals mainly contain the 22 mode, and thus we can see that the GR signal with the transformation $\phi_c \rightarrow \phi_c + \pi/4$ very well mimics the Type-II image, as expected.
	
	We now explore how helpful these distortions in the Type-II image strains can be in separating such signals from a catalog of detected BBH events and, at the same time, how they would affect the inference of parameters if the correct template is not used. Below, we discuss the simulations' details to address these questions.
	
	\begin{figure*}[htb!]
		\centering
		\includegraphics[width=0.5\textwidth]{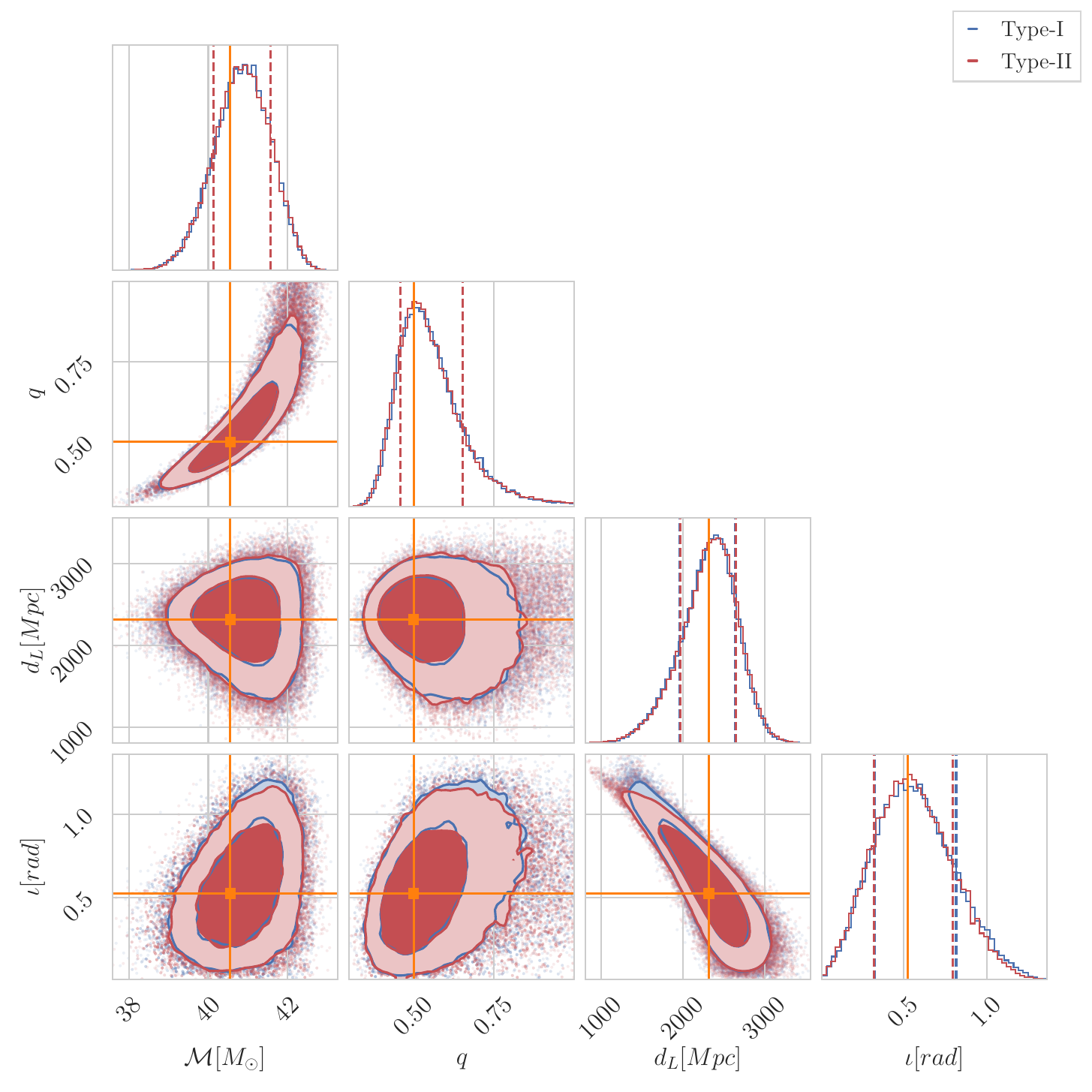}
		\includegraphics[width=0.45\textwidth]{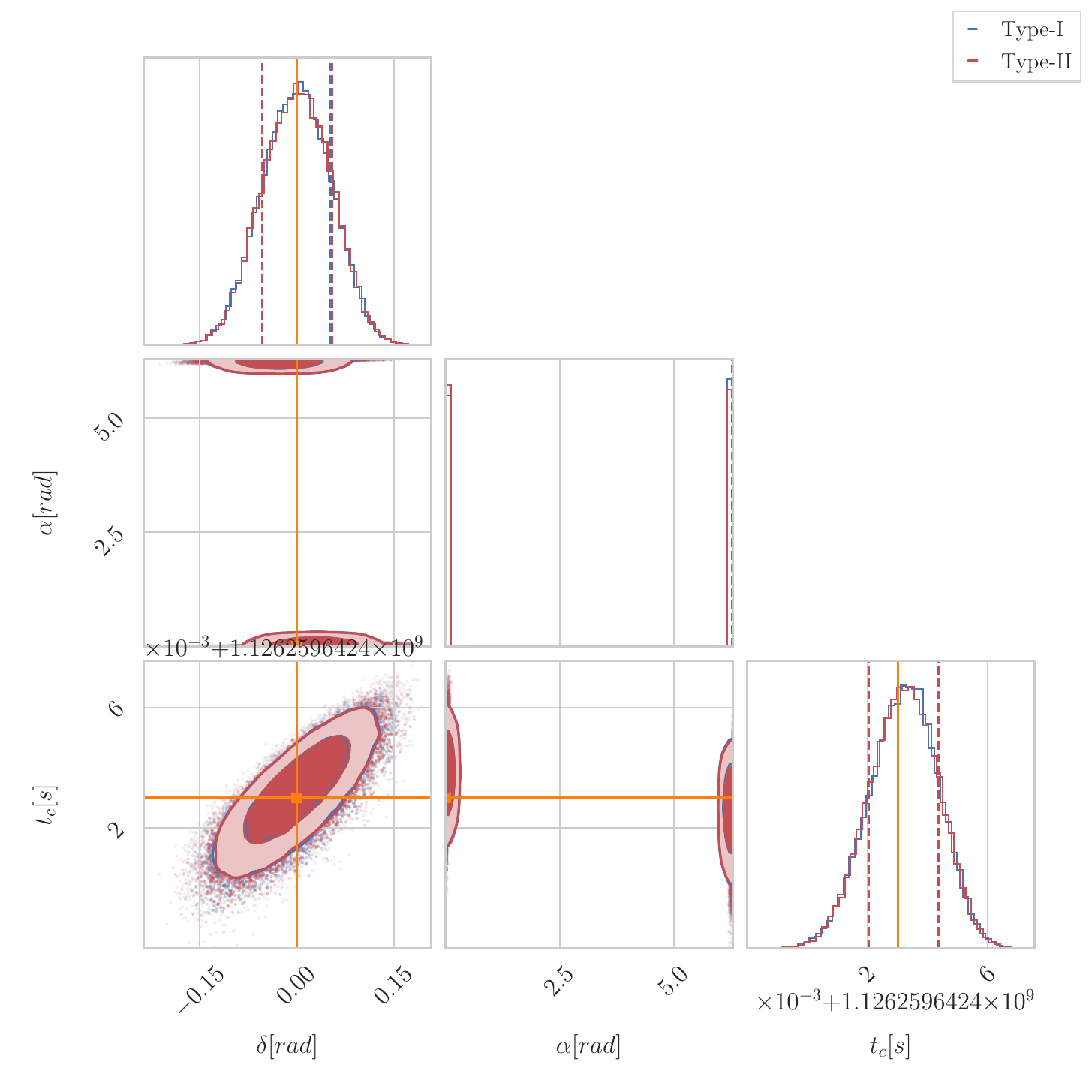}
		\includegraphics[width=0.5\textwidth]{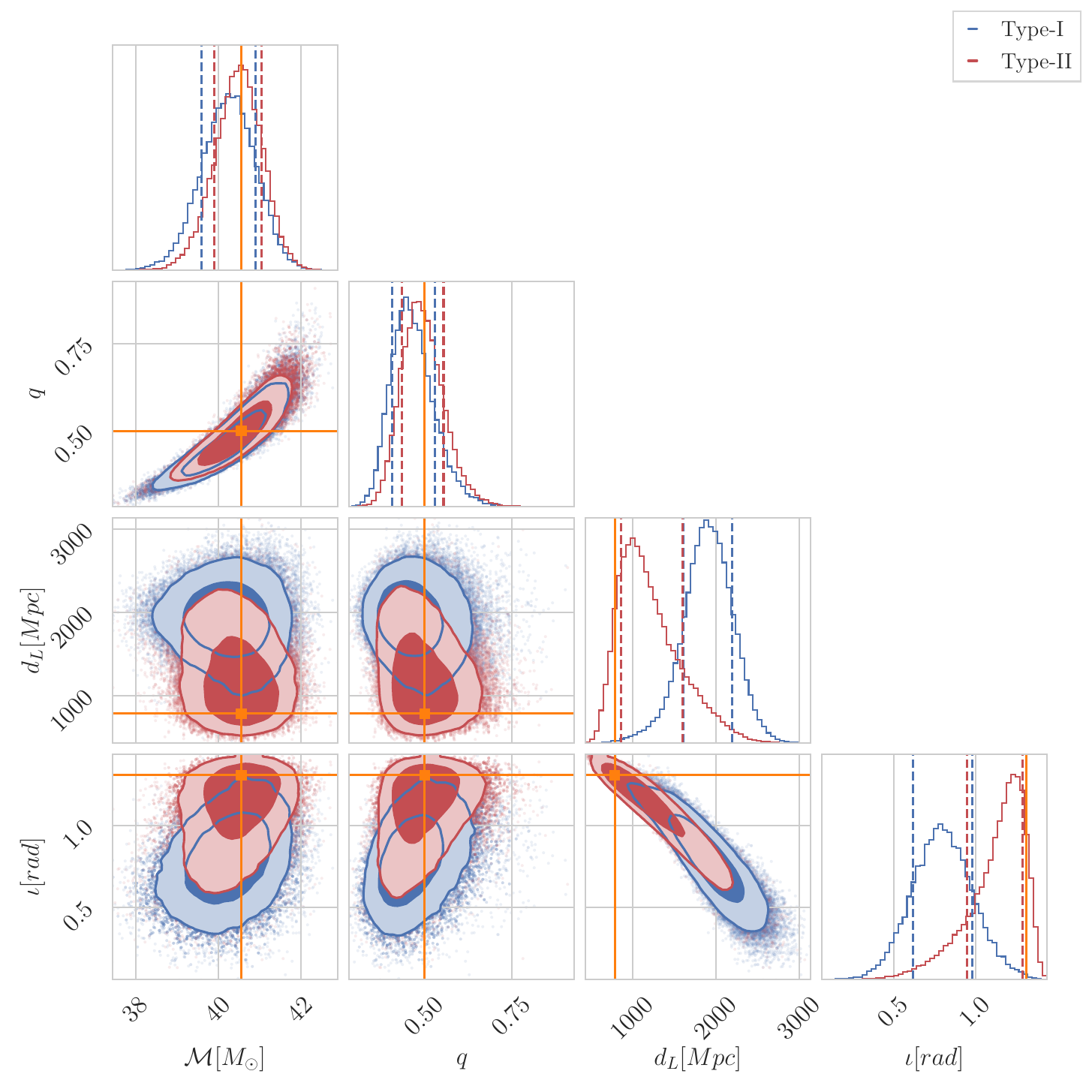}
		\includegraphics[width=0.45\textwidth]{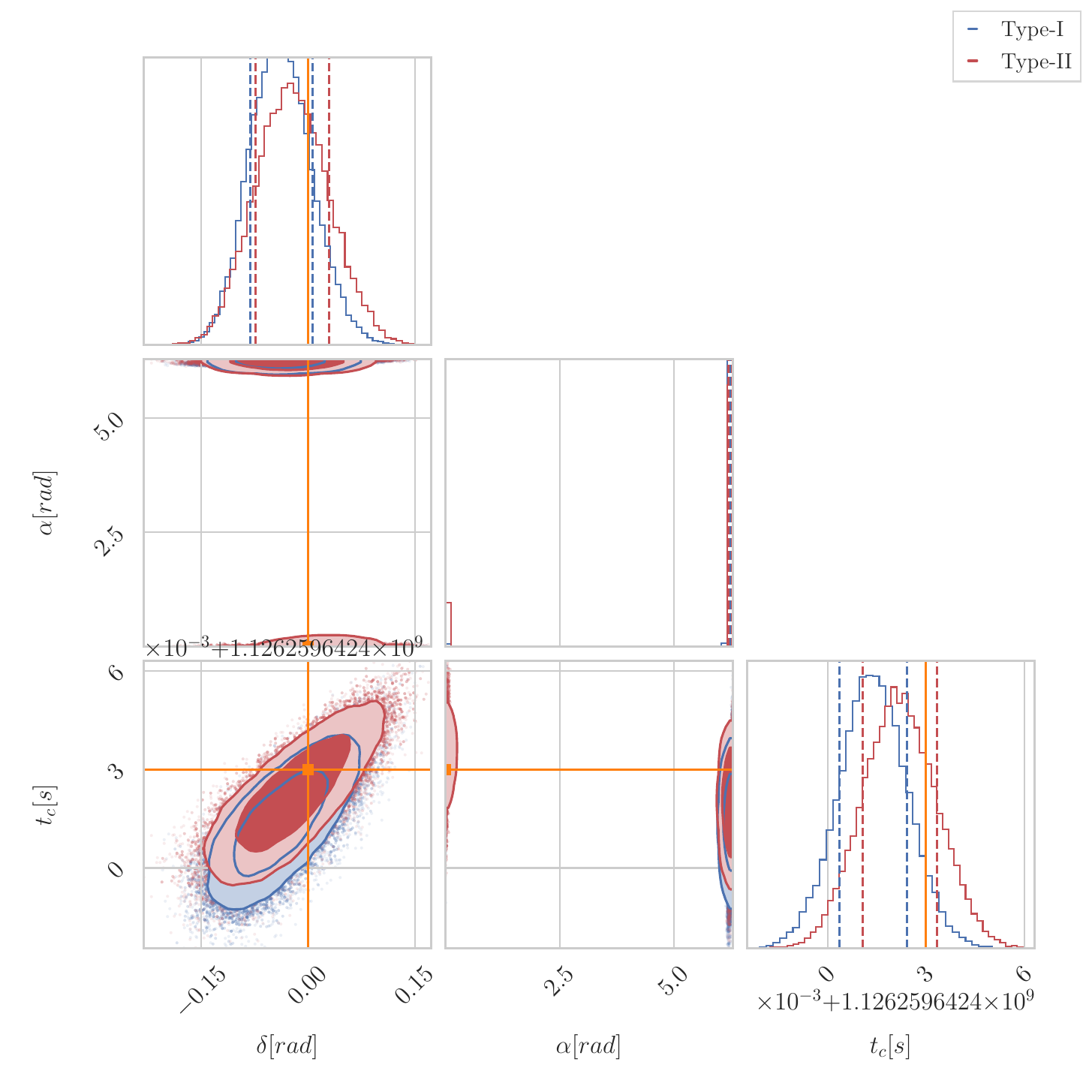}
		
		\caption{ (Top Row) The recovery of source parameters of an injected \TypeII{} signal using \TypeI{} (blue) and \TypeII{} (red) templates for $q=2$, $M=100 M_\odot$, $\rho=20$, and $\iota=\pi/6$. The plots show recovery of chirp mass $\mathcal{M}$, mass ratio $q$, luminosity distance $d_L$, inclination $\iota$, sky locations $\delta$ (dec) and $\alpha$ (RA), and time of arrival $t_c$ in the geocenter frame. The orange lines show the injected value. There is no appreciable difference between the posterior distributions for both the recoveries due to the low inclination and small SNR in the higher modes. (Bottom Row) The recovery of source parameters of an injected \TypeII{} signal using TypeI{} (blue) and \TypeII{} (red) templates for $q=2$, $M=100 M_\odot$, $\rho=20$, and $\iota=5 \pi/12$.  There is now a bias in the inclination and the distance, and the injected value is not recovered within 1$\sigma$. The bias in the distance also biases the recovery of the source frame masses. There are also some insignificant biases in the recovery of the sky locations and time of arrival.}
		\label{fig:corner_plots}
    \end{figure*}

	\begin{figure*}[t]
		\centering
		\includegraphics[width=\textwidth]{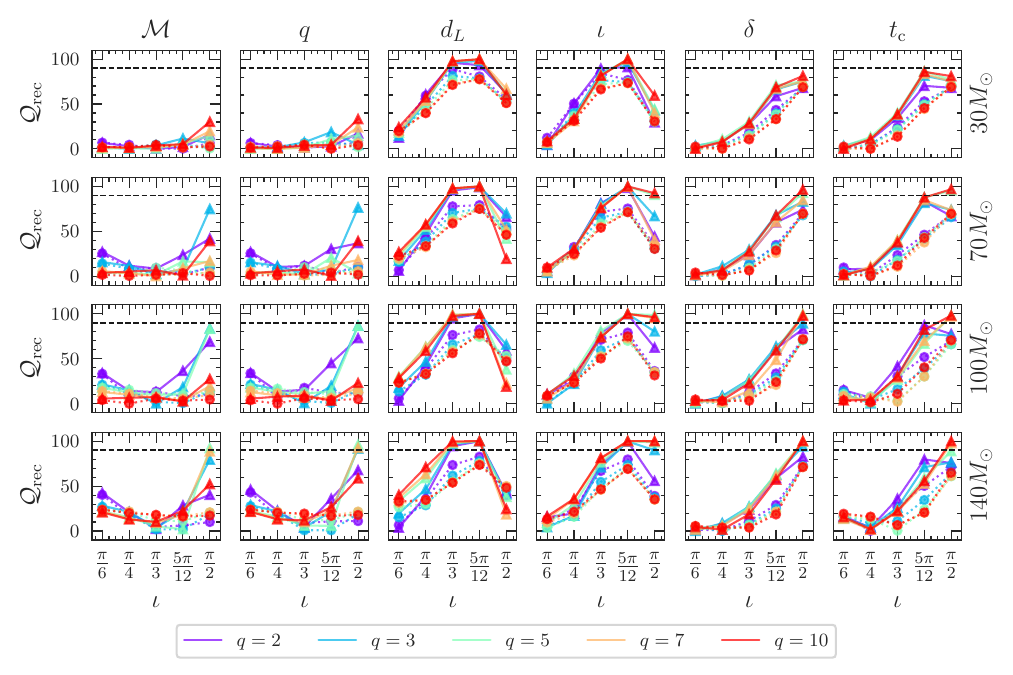}
		\caption{ Each panel shows the quantiles ($\mathcal{Q}_{\mathrm{rec}}$) of the injection value recoveries against the inclination angle ($\iota$) of the injected BBH systems for a given total mass ($M$) (labeled on the right side of the panel) and SNR $\rho=20$. The injections are the Type-II lensed nonspinning BBH signals and the recoveries are with the Type-I template (solid lines) and the Type-II template (dashed lines), accounting for the additional phase shift of $\pi/4$. At the top of the first panel, we label the parameters of the BBH systems that we are interested in. In each plot, the different colors denote the different mass ratios of the BBH systems. A lower quantile value denotes a more accurate inference of a parameter. {The anomalous dips in \qrec{} at high values of $\iota$ are due to bimodalities in the recovery of $t_c$, $\iota$ and $\delta$ at these values. We investigate these further in Appendix~\ref{appendix}.}}
		\label{fig:SNR20_bias}
	\end{figure*}
	
	\begin{figure*}[t]
		\centering
		\includegraphics[width=\textwidth]{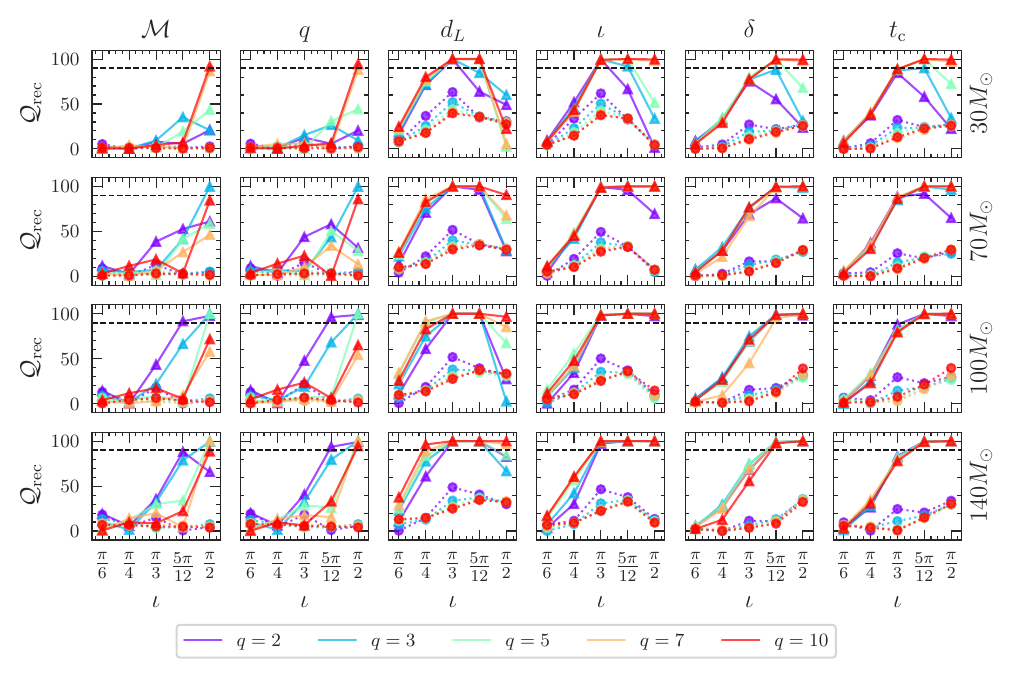}
		\caption{Same as Fig.~\ref{fig:SNR20_bias} but at $\rho = 50$.} \label{fig:SNR50_bias}
	\end{figure*}
	
	\subsection{Simulations and Bayesian inference}
	Measurement of parameters from a GW signal amounts to estimating the posterior probability density function (PDF) $ p (\theta| d) $, where $ \theta $ is the set of parameters that describes a GW signal, and $ d $ is the data stream from the interferometers in the network. According to Bayes theorem, the posterior PDF is given by,
	\begin{equation}\label{key}
	p(\theta | d) = \dfrac{\mathcal{L}(d | \theta) \ \pi(\theta)}{\mathcal{Z}} \qq{.}
	\end{equation}
	Here, $ \mathcal{L}(d | \theta) $ is the likelihood function of the data given the parameters, $ \pi(\theta) $ is the prior distribution for the parameters, and the normalization factor $ Z $ is the evidence, given by 
	\begin{equation}\label{key}
	Z = \int \dd{\theta}\mathcal{L}(d | \theta) \ \pi(\theta) \qq{.}
	\end{equation}
	The evidence can be thought of as the likelihood function marginalized over all the parameters. The evidence is not a useful quantity by itself but is very useful while comparing different models/hypotheses describing the data. In the context of this work, the two hypotheses that we consider are 
	\begin{align*}\label{key}
	&\text{A: the signal is unlensed} \\
	&\text{B: the signal is Type-II lensed}. 
	\end{align*}
	In order to understand which hypothesis fits the data better, we formulate the Bayes factor $ \mathcal{B}^{\text{Type-II}}_{\text{unlensed}} $, which is defined as the ratio of the evidences of the two hypotheses
	\begin{align}
	\mathcal{B}^{\text{Type-II}}_{\text{unlensed}} &= \dfrac{Z_\text{Type-II}}{Z_\text{unlensed}}\\
	\implies \ln \mathcal{B} &= \ln \mathcal{B}^{\text{Type-II}}_{\text{unlensed}} = \ln Z_\text{Type-II} - \ln Z_\text{unlensed}
	\end{align}
	We simulate Type-II image strains using Eq.~\eqref{eq:amp_fact} for total mass $M=m_1+m_2$ $\in \{30, 70, 100, 140\} \msun$, mass ratio $q = m_1/m_2 \in \{2, 3, 5, 7, 10\}$, and inclination $\iota \in \{\pi/6, ~\pi/4, ~\pi/3, ~5\pi/12, ~\pi/2\}$ of the nonspinning BBHs. 
	We scale the luminosity distance of these injections to signal-to-noise ratios (SNRs), $ \rho $, of 20 and 50 in a network with the LIGO-Hanford and LIGO-Livingston detectors at the Advanced LIGO sensitivity, and the Virgo detector at the Advanced Virgo sensitivity \cite{KAGRA:2013rdx}. We choose a realization of the detector noise that is exactly zero at all frequencies (called the ``zero-noise'' realization); this can be thought of as the most probable realization of the noise. This choice allows us to isolate the effects of waveform systematics on parameter estimation from the noise systematics. However, a fully-rigorous analysis would require adding a nonzero noise realization to these injections to make it directly relevant for the LIGO-Virgo observations. We leave this for future studies. Nevertheless, the SNRs (such as 50) considered in our work are expected to be high enough to minimize the effect of noise systematics in our results.
	
	We perform full Bayesian parameter estimation on these injections (simulations) using Type-I (Eq.~\eqref{eq:final_strain_type1}) and Type-II (Eq.~\eqref{eq:final_strain_type2}) templates, keeping the priors on the source parameters the same between the recoveries for each injection. The priors we choose are uniform in the detector-frame chirp mass~\footnote{The chirp mass $\mathcal{M}$ is a mass parameter which describes the inspiral part of the BBH signals at leading order.} and the mass ratio of the binaries, uniform in comoving volume for the luminosity distance, and uniform in sky location and inclination of the binaries. The parameter estimation runs are performed using \textsc{bilby} and \textsc{bilby\_pipe} \cite{Ashton:2018jfp,Romero-Shaw:2020owr} coupled with the dynamic nested sampler \textsc{dynesty} \cite{Speagle:2019ivv}. We use \textsc{IMRPhenomXHM} \cite{Pratten:2020ceb} as our waveform approximant for the runs; this approximant contains the 33, 44, 21 and 32 modes of the gravitational wave signal in addition to the dominant 22 mode. 
	
	\section{Results}
	\label{sec:results}
	
    Fig.~\ref{fig:Bayes_Factor} shows the Bayes factor between the Type-II and Type-I recoveries of Type-II injected BBH signals at the $\rho=20$ and 50, respectively, for various total masses. We choose a threshold of $\ln \mathcal{B}=2$ for the distinguishability of the Type-II recovery against the Type-I recovery of Type-II injected signals \cite{Jeffreys:1939xee, Kass:1995loi}. {Choosing a higher threshold for the distinguishability would require the lensed events to be observed at higher SNRs keeping other parameters fixed}. The triangle shapes in  Fig.~\ref{fig:Bayes_Factor} denote the signals which pass this threshold. We can see that at the $\rho=20$, independently of the total mass and the mass ratio, when the inclination $\iota \gtrsim 5\pi/12$, the Type-II image signals become distinguishable from the Type-I (or the unlensed signals). {At higher SNRs, e.g., $\rho=50$, even less inclined binaries ($\iota \sim \pi/3$) will allow us to distinguish the lensed signals from the unlensed ones. This is because, at higher SNRs, the effect of HMs is more significant}. Since $\ln \mathcal{B} \sim \rho^2$, we expect that at even higher SNRs (e.g., 100 or larger), even smaller inclinations ($\sim \pi/4$) should allow us to distinguish the Type-II images. This means that third-generation detector networks, where SNRs of $\sim 100$ could be typical, would easily allow us to distinguish the Type-II images even when the inclination is low. This prediction is also consistent with results from Ref. \cite{Wang:2021kzt}. Although higher inclinations permit better measurement of Type-II imaegs, they are intrinsically less detectable; we note that $\iota \gtrsim 5\pi/12, \pi/3, \pi/4$ make up $\sim 5\%, 14\%, 34\%$ of the total detectable (lensed or unlensed) population of sources respectively, based on the detected distribution of inclinations $p_\mathrm{det}(\iota)\sim (1 + 6 \cos^2\iota + \cos^4\iota)^{3/2} \sin \iota$~\cite{Schutz:2011tw}.

    For the distinguishable Type-II image signals, we then explore the effect on the recovery of source parameters if they are recovered with Type-I image (or equivalently, unlensed) templates. The top row in Fig.~\ref{fig:corner_plots} compares the recovery of an injected Type-II image having $\rho=20$, $ M=100 M_{\odot} $, $ q=2 $, and $ \iota=\pi/6 $ with both Type-I and Type-II templates. We see that the injected values lie well within the posterior PDFs for both recoveries, and there are no discernible differences between the two recoveries. However, increasing $ \iota $ to $ 5 \pi / 12 $ (bottom row of Fig.~\ref{fig:corner_plots}) keeping the other parameters fixed causes the injected masses, distance and inclination values to lie in a region with low posterior probability if the recovery is made with a Type-I template. This shows that the distortions in the waveform can cause significant biases in parameter estimation in certain regions of the parameter space. We investigate this further in Fig.~\ref{fig:SNR20_bias} and \ref{fig:SNR50_bias}, where we show results of recoveries with Type-I image templates (Eq.~\eqref{eq:final_strain_type1}) denoted by solid lines and with Type-II image templates (Eq.~\eqref{eq:final_strain_type2}) denoted by the dashed lines at SNR of 20 and 50, respectively, for different total masses of the Type-II injected signals. On the y-axis, we show the quantiles \qrec{} of the recovery of the injected values for the significantly affected parameters. For example, a $\qrec=90\%$ denotes that the injected value of the parameter is recovered at a value that forms one edge of the $ 90\% $ posterior area; in other words, the injection value is recovered in the tail of the posterior. Such a case would represent a Type-II signal which would provide a biased estimate of this parameter upon recovery with the unlensed template. 
	
	We can see from Fig.~\ref{fig:SNR20_bias} that, at $\rho=20$, for the Type-II signals with lower total mass ($M=30M_{\odot}$), the intrinsic parameters chirp mass $\mathcal{M}$ and mass ratio $q$ are recovered at nearly the exact quantiles, i.e., the phase shift does not affect their recoveries. This could be because, as explained before, the HMs are not very relevant for the inspiral-dominated signals at low SNRs. At higher SNRs, however, the difference between the quantiles of their recoveries becomes significant at high inclination angles ($\iota\gtrsim 5 \pi/12$) (Fig.~\ref{fig:SNR50_bias}). At such high SNRs, for higher total mass systems, the differences become significant even at the inclination $\iota\sim \pi/3$. This {value of} the inclination would {further} decrease with increasing SNR. We thus expect that the chirp mass and mass ratio would be affected significantly for the 3G detectors when $\rho \sim 100$ or greater could be achievable. We note that even at $\rho=50$, there are multiple binaries for which the quantiles of the chirp mass and mass ratio recoveries exceed 90\%, and thus their estimates are completely biased (see Fig.~\ref{fig:SNR50_bias}).
	
	{The inference of the distance ($d_L$) and the inclination also get affected significantly at higher SNRs. At lower SNR (Fig.~\ref{fig:SNR20_bias}), there is not much difference between their recovery quantiles. However, the quantiles get pushed beyond 90\% for a range of inclination angles ($\pi/3 \lesssim \iota \lesssim 5 \pi/12$). At higher SNR (Fig.~\ref{fig:SNR50_bias}), on the other hand, the binaries even with the inclination angle of $\sim \pi/4$ face significant differences. We also find that the declination ($\delta$) and coalescence time ($t_c$) are affected significantly at higher SNR. We can see from Fig.~\ref{fig:SNR50_bias} that higher total mass systems with the inclination angle of $\sim \pi/3$ onward have completely biased estimates of these parameters. Again, increasing the SNRs further would lower the inclination at which parameters get biased further. We thus expect that most of the Type-II lensed events in the 3G era will undergo biased estimates of these quantities, most notably the sky location. }
	
	{We note that} there are some seemingly anomalous features at $ \iota=\pi/2 $ in Fig. \ref{fig:SNR20_bias} and Fig. \ref{fig:SNR50_bias}, where the value of $  \mathcal{Q}_\mathrm{rec}$ drops as compared to $ \iota = 5 \pi/12 $. These occur due to bimodalities in $ t_c $, $ \delta $ and $ \iota $ posteriors (see Appendix \ref{appendix} for more details), thus rendering $ \mathcal{Q}_\mathrm{rec}$ to be an insufficient quantifier for the bias at $ \iota = \pi/2 $. 
	
	\section{Conclusion and Future Work}
	\label{sec:conclusion}
	
	Strong lensing produces three types of images; Type-I, Type-II and Type-III, corresponding to the minima, saddle and maxima of the total arrival time of the lensed GWs. While Type-I and III images do not affect the frequency profiles of the unlensed BBH signals, the Type-II image can, depending on the morphology of the signals. {For example, if the unlensed BBH signals have support for HMs, precession, eccentricity or any combination of these, the Type-II image strains would receive additional distortions due to the phase shift ($\pi/2$) caused by strong lensing. In this work, we explored the possibility of identifying Type-II lensed signals usingthese distortions in the observed signals.}
	
	Using a set of full parameter estimation recoveries for \TypeII{} lensed injections on a varied parameter space, we showed that it would be possible to ascertain the \TypeII{} nature of the most lensed BBH signals at high SNRs. This becomes very relevant for the 3G detectors and beyond, where such high SNR events could be frequently expected. This study implies that we should be able to tell if an event is \TypeII{} lensed or not just by the observation of a single image. 
	
	We then showed that the identifiable Type-II images, which are of our interest here, will have sufficient distortions caused by the lensing phase shift $\pi/2$ that when they are recovered with Type-I image templates, would recover significantly biased parameters of the underlying BBH signals. We quantified these biases for a range of inclinations, total mass, and mass ratios of the binary. Our results suggest that the recovery of parameters such as the chirp mass, mass ratio, coalescence time, luminosity distance, inclination and the sky position of the binary becomes significantly biased. Since the first-cut search for lensed pairs involves demanding consistency between the sky positions of the images, we suggest that parameter estimation should be performed using Type-II templates on all signals {because, \textit{a priori}, we would not know the nature of the image types as well as the parameters of the signals}.
	
    In this work, we have considered  nonspinning quasicircular BBH signals to concentrate on the effects of only including higher harmonics of the radiation. Extending this study to generic spinning BBH signals with eccentricity would be natural. We, however, expect that precession and/or eccentricity would cause similar biases~\cite{Ezquiaga:2020gdt}. We have already verified this from a few precessing BBH Type-II injections, but a thorough study will shed more light on this. We also have not considered the impact of noise systematics on our ability to identify Type-II lensed signals. We leave these investigations for future work, along with an exploration of the detectability of these effects for a poulation of simulated sources..

	\section{Acknowledgements}
	We thank Jose Maria Ezquiaga, Justin Janquart and Otto Hannuksela for discussions. We also thank Bala Iyer, Parameswaran Ajith and other members of the Astrophysical Relativity group at ICTS for feedback, and Sanskriti Chitransh for a careful reading of the draft. Computations were performed on the Alice cluster at ICTS. AV's research is supported by the Department of Atomic Energy, Government of India, under Project No. RTI4001. This work makes use of \textsc{NumPy} \cite{vanderWalt:2011bqk}, \textsc{SciPy} \cite{Virtanen:2019joe}, \textsc{Matplotlib} \cite{Hunter:2007}, \textsc{jupyter} \cite{jupyter}, \textsc{dynesty} \cite{Speagle:2019ivv}, \textsc{bilby} \cite{Ashton:2018jfp}, \textsc{PyCBC} \cite{pycbc}, \textsc{LALSuite} \cite{lalsuite} and \textsc{PESummary} \cite{Hoy:2020vys} software packages.

    \appendix
    \section{Investigating patterns in the bias}
\label{appendix}
\begin{figure*}[htb!]
	\centering
	\includegraphics[width=0.5\textwidth]{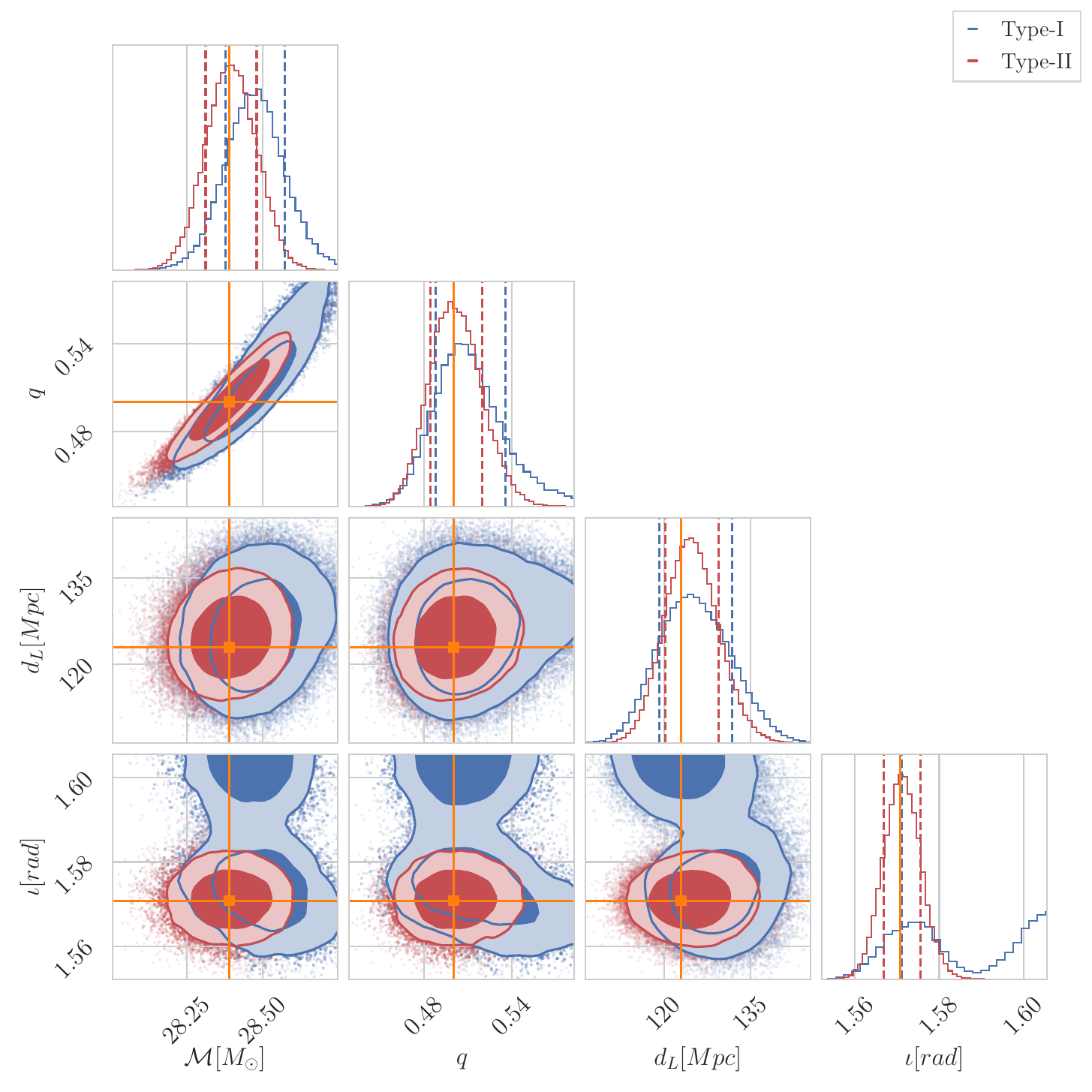}
	\includegraphics[width=0.45\textwidth]{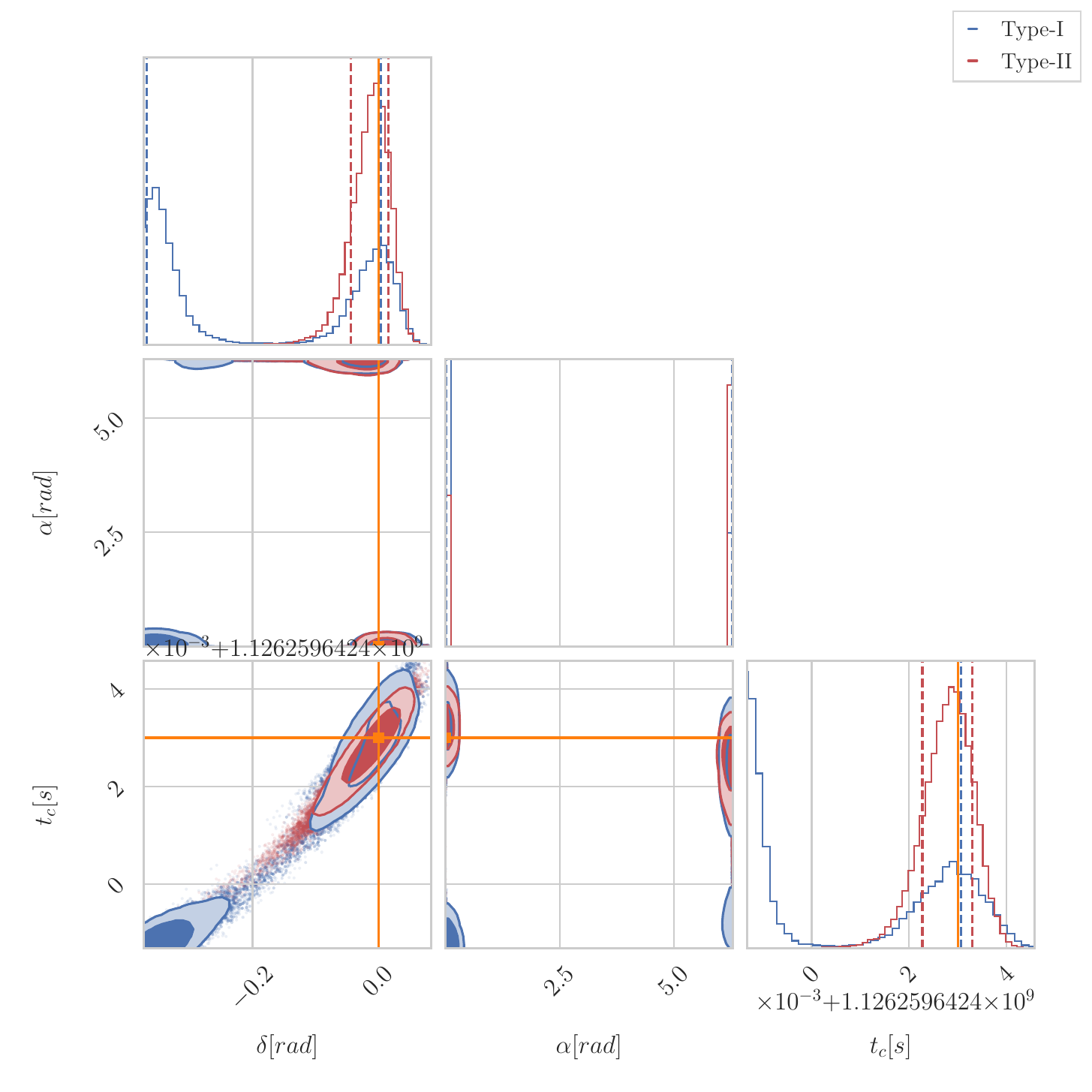}
	\caption{The recovery of source parameters using \TypeI{} (blue) and \TypeII{} (red) templates for $q=2$, $M=70 M_\odot$, $\rho=50$, and $\iota=\pi/2$. The \TypeII{} recoveries are consistent with the injection values, but the \TypeI{} recoveries have correlated bimodalities $ t_c $, $ \delta $ and $ \iota $. One of these modes is consistent with the injected values, while the other mode is far away from the injected values. The distance posterior does not have a significant bias with respect to the injected value; however, the width of the posterior is larger than that of posterior in the \TypeII{} case.}
	\label{fig:q05_inc90_mtot70}
\end{figure*}

\begin{figure*}[htb!]
	\centering
	\includegraphics[width=\textwidth]{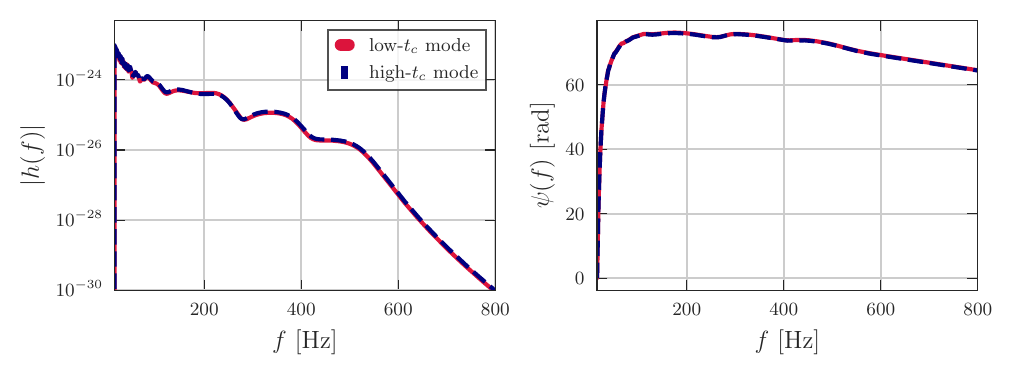}
	\caption{Comparison of the waveform amplitude $ |h(f)| $ and phase $ \psi(f) $ (projected onto the LIGO-Hanford detector) for the maximum likelihood samples from the low-$ t_c $ and high-$ t_c $ modes. There is considerable agreement between the two waveforms.}
	\label{fig:fdomain_amplitude_phase}
\end{figure*}

For some cases in Fig. \ref{fig:SNR20_bias} and Fig. \ref{fig:SNR50_bias}, there are anomalous dips in the bias (quantified by $\mathcal{Q}_\mathrm{rec}$) at $\iota= \pi/2$. We investigate these below.

In order to illustrate the causes of these dips, we show the corner plot of source parameter recovery for a Type-II injection with $ q=2 $, $ M=70 M_\odot $, $ \rho=50 $ and $ \iota = \pi/2 $ with both Type-I and Type-II templates (Fig. \ref{fig:q05_inc90_mtot70}). The posteriors obtained with the Type-II template do not show any anomalous features, and the injection values lie near the peak of the posterior. However, this is not the case for Type-I recovery. The first thing to note is that the $ t_c $ posterior is bimodal, thus also causing the $ \delta $ posterior to be bimodal since there is a correlation between the two parameters. This is perhaps not too surprising since, as we have shown earlier, Type-II effects distort the waveform away from GR considerably. One of the modes in the $ t_c - \delta $ plane includes the injected value, while the other is far away from it. This bimodality also effectively broadens the region where the posterior has significant support, thus causing the $ \mathcal{Q}_\mathrm{rec} $ to be shifted to lower values. This is the reason for the anomalous behavior seen at $ \iota = \pi/2 $, and one can also see that a similar argument holds for the recovery of $ \iota $ as well. On the other hand, the $ d_L $ posterior peaks at the injected value, but the posterior width is larger than that in the Type-I recovery case. This, again, causes a decrease in the value of $ \mathcal{Q}_\mathrm{rec} $ estimated for the $ d_L $ parameter. The points noted above show that $ \mathcal{Q}_\mathrm{rec} $ is not a good enough quantifier of the bias for these edge cases.

We have also checked that the likelihood values for samples at both modes are very similar. To check whether the waveforms themselves at these two modes are similar, and to ensure sanity of the likelihood calculation and the sampling, we plot the frequency-domain amplitude and phase of the maximum likelihood waveforms (projected onto the LIGO-Hanford detector) from both the modes (Fig. \ref{fig:fdomain_amplitude_phase}). As one can clearly see, the waveforms match very well in the frequency domain. 

	\bibliography{references}
	
\end{document}